\documentstyle[prl,aps,epsf]{revtex} \def\narrowtext{} \tighten \twocolumn
\input epsf.sty
\begin{document}
\draft

\title{Magnetic Resonance, Electronic spectra and Bilayer
Splitting in Underdoped Bi2212}
\author{
        J. Mesot,$^1$
        M. Boehm,$^{1,2}$
        M. R. Norman,$^3$
        M. Randeria,$^4$
        N. Metoki,$^{5}$
        A. Kaminski,$^6$
        S. Rosenkranz,$^6$
        A. Hiess,$^2$
        H. M. Fretwell,$^7$
        J. C. Campuzano,$^{6,3}$
 and K. Kadowaki$^8$
       }
\address{
         (1) Laboratory for Neutron Scattering, ETH Zurich and PSI
             Villigen, CH-5232 Villigen PSI, Switzerland\\
         (2) Institute Laue-Langevin, BP 156, F-38042 Grenoble Cedex\\
         (3) Materials Sciences Division, Argonne National Laboratory,
             Argonne, IL 60439\\
         (4) Tata Institute of Fundamental Research, Mumbai 400005,
             India\\
         (5) Advanced Science Res. Center, Japan Atomic Energy Res. Inst.,
             Tokai, Ibaraki 319-1195, Japan\\
         (6) Department of Physics, University of Illinois at Chicago,
             Chicago, IL 60607\\
         (7) Department of Physics,University of Wales,
             Swansea, UK\\
         (8) Institute of Materials Science, University of Tsukuba,
             Ibaraki 305, Japan\\
         }
\address{%
\begin{minipage}[t]{6.0in}
\begin{abstract}
We report the first inelastic neutron scattering (INS) experiments of the
resonance in the 
underdoped regime of Bi$_{2}$Sr$_{2}$CaCu$_{2}$O$_{8+\delta}$. The
energy of the resonance is found to be 34 meV and the temperature
dependence shows a smooth evolution through $T_{c}$ with a
remnant persisting in the pseudogap state. Besides the INS data,
we present also angle resolved photoemission (ARPES)
spectra taken on the same crystals. As a function of temperature,
the neutron intensity scales with the square of the ARPES gap estimated
by the leading-edge midpoint at $(\pi,0)$.
We also show that the energy of the collective mode inferred from
the peak-dip-hump structure compares well with the energy of the resonance.
Based on kinematics, we argue that the correlation of these two energies
is consistent with the absence of bilayer splitting in the electronic
dispersion.
\typeout{polish abstract}
\end{abstract}
\end{minipage}
}

\maketitle
\narrowtext

It is now widely accepted that strong electronic correlations are
responsible for many unusual properties of the high
temperature cuprate superconductors (HTSC)\cite{ANDERSON92}.
Evidence is 
provided by momentum resolved spectroscopies such as
inelastic neutron scattering (INS) and angle resolved photoemission
spectroscopy (ARPES), which probe the fundamental excitations connected
with the spin and single particle response functions.

In the normal state of underdoped HTSC, no quasiparticles can be
observed \cite{JC99} suggesting unusually strong electronic correlations.
The situation changes drastically below $T_{c}$, since well defined
quasiparticles 
can be observed at each point of the Fermi surface\cite{KAMINSKI00}.
In particular at the $(\pi,0)$ point, the low-temperature spectral function
is 
characterized by a peak-dip-hump structure which is believed to be due
to the interaction of the electrons with an electronic collective
mode\cite{NORMANPRB98}.

Strong correlations are also inferred from INS since, unlike
in weakly correlated metals, spin excitations could be
measured in La$_{2-x}$Sr$_{x}$CuO$_{4}$\cite{KASTNER98},
YBa$_{2}$Cu$_{3}$O$_{x}$\cite{BOURGES98,OPYBCO,UDYBCO} (YBCO),
and recently in
optimally and overdoped
Bi$_{2}$Sr$_{2}$CaCu$_{2}$O$_{8+\delta}$\cite{FONG99}
(Bi2212).  While for the optimally and overdoped YBCO and Bi2212 samples,
a sharp resonance can be observed only below the critical temperature
$T_{c}$,
data obtained in the underdoped regime of
YBCO suggest that a broadened remnant persists above $T_{c}$.

The question of the origin of the resonance has been widely addressed and
several models 
involving a spin exciton in the particle-hole\cite{PH}
channel, a collective mode in the
particle-particle\cite{PP} channel, or antiferromagnetic domains due to
formation of stripes\cite{EMERY93} have been
proposed. In order to establish whether or not the magnetic resonance
is relevant to the origin of high temperature superconductivity,
quantitative comparisons
of experimental results obtained by different techniques are very
useful. An example 
is given by the attempt to compare the weight of the resonance mode
and the superconducting condensation energy in YBCO\cite{DAI99}.
In a related vein, we have recently shown that the doping dependence of
the collective mode energy inferred from ARPES in
Bi2212 compares favorably with the resonance's energy
measured by INS in YBCO\cite{JC99}.
So far, a direct comparison between ARPES and INS measurements in the
underdoped 
regime of Bi2212 could not be realized since all INS data
in that doping range have been obtained only on YBCO.

We report in this paper the first INS measurement of the resonance in
underdoped ($T_{c}=70K$) Bi2212. In parallel, we have performed
temperature-dependent ARPES measurements of the spectral function at the
($\pi$,0) point on the \textit{same sample}.

The INS measurements were performed on the triple-axes spectrometer IN8 at
the 
Institut Laue-Langevin in Grenoble, France. We used a vertically focused
Cu(111)-monochromator and
a vertically and horizontally focused PG(002)-analyser with 35 meV fixed
final energy. The wave vector $\mathbf{Q}$ is given in units of
$(1/d_{\parallel},1/d_{\parallel},1/d_{\perp})$, where
$d_{\parallel}$ is the Cu-Cu in-plane distances and
$d_{\perp}$ is the distance between two adjacent Cu-O planes.

The sample was built of an alignment of five crystals with an overall
volume of about 80 $mm^{3}$ and overall mosaicity $< 3^{o}$. All
crystals were grown from the same batch and showed a critical temperature
$T_{c}=70K$ and a transition width of 2-4K (see insert of Fig.~1).

Fig.~1a shows energy scans at $\textbf{Q}=(\pi,\pi,-3\pi)$ taken at
$T=10K$ and $T=250K$. As it is known from phonon density-of-state
measurements\cite{PINTSCHOVIUS98},
the spectra, for energies $20< E < $40 meV,
are contaminated by phonon branches whose intensities, following Bose
population factors $B(T,E)=(1-\exp{(-E/k_{B}T)})^{-1}$,
will increase with increasing temperature $T$. In order to separate out
the lattice contribution from the magnetic one, and knowing that phonons
are strongest around 20 meV, we normalize the
data at 24 meV, after subtraction of a flat background\cite{BOSE}.
We observe that at low temperature, a clear signal
emerges between 30 and 40 meV. A fit to the difference
of the low- and high-temperature data shows that the peak is
centered at 34$\pm$2 meV (Fig.~1b)\cite{NOTE1}. This value agrees with the
energy of the resonance measured in underdoped YBCO of similar doping
level\cite{UDYBCO}.
The magnetic origin of the peak is further confirmed by the observation
of the characteristic modulations of the
resonance's intensity along the in- (Fig.~2b) and out-of-plane directions
(Fig.~2a), confirming its $(\pi,\pi,\pi)$ (odd) symmetry\cite{BOURGES98}.
The odd symmetry along the z-direction will turn out to be crucial
for an understanding of the ARPES spectra. The in-plane Q-width of
the resonance is 0.59 $\pm$ 0.07 $\AA^{-1}$ at low temperature, a value
comparable to that measured for optimally doped Bi2212\cite{FONG99}
but about twice as large as the low-temperature value measured in
underdoped YBCO\cite{BOURGESSCIENCE00}.

As the temperature is increased, the energy-integrated weight of the
neutron signal (see Fig.~3) is found to weaken regularly,  giving
rise to a remnant far into the pseudogap state, similar to what has
been observed in underdoped YBCO\cite{BOURGES98,UDYBCO}. Notice
that no break at $T_{c}$ can be resolved.

We now turn to the ARPES data obtained on three portions
of the crystals used for the INS experiments.
The experiments were performed at the Synchrotron Radiation Center,
Wisconsin, using a high-resolution 4-meter normal incidence monochromator,
with a resolving power of $10^{4}$ at $10^{11}$
photons/sec. We used 22 eV photons, with a 20 meV (FWHM) energy
resolution, and a momentum window of radius 0.045$\pi$.

Fig.~4a shows ARPES data taken at $(\pi,0)$ as a function of temperature.
The low temperature spectra consist of a sharp spectral peak,
followed at
higher binding energy by a pronounced spectral dip, then by a broad maximum
(the hump).  This structure evolves with temperature into a single broad
peak with a leading edge gap (the pseudogap).
For binding energies above the hump energy,
the spectra are temperature independent.
Below, we will comment further
on the temperature evolution, but for now, we will try to make a correlation
of the energy scales apparent in the low temperature ARPES spectra with the
INS resonance energy.
Fig.~4b shows an expanded view of the near-E$_{F}$ region and we find that
the peak- (=$\Delta_{0}$), dip- and hump-energies are 48$\pm$ 2, 80$\pm$ 2
and
193$\pm$ 5 meV, respectively, characteristic of underdoped Bi2212
samples\cite{JC99}. $\Delta_{0}$ represents the maximum of the d-wave
superconducting gap.

It has been proposed that the peak-dip-hump structure is caused by
the interaction of the electrons with a collective
mode whose energy $\Omega<2\Delta_{0}$\cite{NORMANPRB98}.
This can be understood in the following way:
for quasi-two dimensional systems and within the impulse
approximation, ARPES measures the \textbf{k} and $\omega$ dependence of the
single particle spectral function\cite{RANDERIA95}.
For binding energies $\omega<\Delta_{0}+\Omega$, the inelastic scattering
process by the collective mode
is forbidden since the final state of the electrons would lie in the gap.
As a consequence the lifetime of the quasiparticles is
long which gives rise to the sharp peak in the ARPES spectra. For
binding energies $\omega>\Delta_{0}+\Omega$, a relaxation channel opens and
the
lifetime drops, creating a spectral dip. The dip is further enhanced by a
maximum of $\Sigma^{\prime}$ at
$\omega=\Delta_{0}+\Omega$ which is required by the Kramers-Kronig relations
for the self-energy \cite{NORMANPRB98}. It is then easy to
realize that the energy of the mode can be directly inferred from the
ARPES data by measuring the energy difference between the dip and the
peak. For this underdoped sample we obtain 32
$\pm$ 4 meV for the mode energy,
which is in excellent agreement with the value of 34 $\pm$ 2 meV obtained
for the resonance from INS (see Fig.~1b). In order to illustrate this
correspondence between the ARPES collective mode and the INS resonance,
we have also drawn, in Fig.~4b, the magnetic resonance
(bold-broken line), with its energy-axis shifted by 48 meV, the
peak energy. 

We now return to the evolution of the ARPES lineshape as a function of
temperature (Fig.~4a).
As the temperature is raised, the quasiparticle peak smears out, and
the spectral dip fills in, leading eventually to a single broad peak with
a leading edge gap\cite{FOOTNOTE}. Fig.~3b shows (open
symbols) the temperature evolution of the square of the gap estimated by
the leading-edge midpoint
(normalized to the $T=50~K$ value). Below $T_{c}$, it
remains constant and then decreases substantially above $T_{c}$.
This temperature evolution is very similar to that of the neutron intensity,
which we also plot (closed symbols).  The neutron data points were obtained
from Fig.~3a by subtracting the $T=250~K$ value.

We remark that the correlation we see in Fig.~3b was first suggested by
Demler and Zhang based on a particle-particle interpretation for the
neutron resonance\cite{PP}.  In such a picture, the neutron resonance
energy should not depend on temperature (which we also observe), but its
intensity should scale with the square of superconducting energy gap.
This scaling has been verified for the optimally- to over- doped region of
the phase diagram of YBCO\cite{OPYBCO} and Bi2212\cite{FONG99}.
For our underdoped sample, it is interesting to note that,
within the error bars,
we also find that this scaling persists in the pseudogap phase.
This is strong evidence that the leading edge pseudogap is a consequence of
pairing
correlations\cite{RANDERIA98}.  A similar conclusion has been reached
recently 
by a neutron scattering study of the field dependence of the
resonance\cite{DAI00}.

We finally discuss the implication of our results for the question
of bilayer splitting below $T_c$ \cite{DING96}.
On general grounds we expect that
the two CuO$_{2}$ layers hybridize to form non-degenerate bonding (B) and
antibonding (A) states.
We now argue that, for the observed \cite{FRETWELL00} hole-like Fermi
surface, only a small and most likely vanishing bilayer
splitting between the A and B states is consistent with the kinematics
shown in Fig.~5. If the two states were split the quasiparticle
peaks for the two would be at the corresponding gaps $\Delta_A$ and
$\Delta_B$.
Because of the $(\pi,\pi,\pi)$ symmetry of the resonance,
scattering by the resonance implies both in- and out-of-plane
wavevector transfers. The in-plane component connects the
$(\pi,0)$ and $(0,\pi)$ points while the $Q_z = \pi$ out-of-plane component
connects bands A and B as shown in Fig.~5.
Thus the threshold for decay, i.e., the dip energy, at
${\bf k} = (\pi,0)$ via the mode of energy $\Omega$
is at $\Delta_B + \Omega$ for the A state, and $\Delta_A + \Omega$ for the
B state. 
It follows that the ARPES spectra consist of a weighted superposition of
bands 
A and B, and that the measured peak and dip positions are averaged values
of the positions in both channels.
If there were significant bilayer splitting,
the peak would be broadened (or even duplicated) and the dip less sharply
defined. 
Furthermore, the averaged peak-dip difference is equal to
$\Omega$+bilayer splitting. Since we obtained 32 $\pm$ 4 meV and 34 $\pm$ 2
meV 
for the peak-dip and INS resonance energies, respectively, we can put an
upper bound of 8 meV for the bilayer
splitting of the underdoped sample studied here.
This is in agreement with all
known ARPES experiments
on Bi2212 since, even with resolutions as high as 10 meV, no
bilayer splitting could be detected\cite{KAMINSKI01}.

In conclusion we have shown, performing both INS and ARPES experiments
on the same crystals, that the energy of the mode inferred from the ARPES
spectra agrees well with the energy of the INS magnetic resonance.
It is also found that a magnetic remnant persists well into the pseudogap
state, whose intensity as a function of temperature scales with the
leading-edge gap measured at $(\pi,0)$ from ARPES.
We have also argued that the correspondence of the ARPES
peak-dip separation with the INS resonance energy implies that the bilayer
splitting in the electronic dispersion must be very small.

This work was supported by the Swiss National Science Foundation,
the U. S. Dept. of Energy,
Basic Energy Sciences, under contract W-31-109-ENG-38, the National
Science Foundation, through the grants DMR 9624048 and PHY94-07194,
the CREST of JST, and the Ministry of Education, Science,
and Culture of Japan. MR is supported by the Swarnajayanti fellowship
of the Indian DST.

\begin{figure}
\epsfxsize=3.0in
\epsfbox{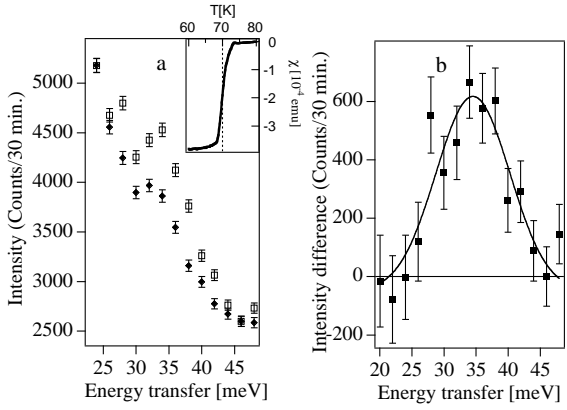}
\vspace{0.5cm}
\caption{a) Energy spectra of the neutron intensities at wave vector
$\textbf{Q}=(\pi,\pi,-3\pi)$ at temperatures $T=10K$ (open squares) and
$T=250K$ (filled diamonds). The
spectra have been normalized according to the procedure described in
the text. Insert: magnetization curve of one of the Bi2212-crystals
used in both the INS and ARPES experiments.
The dashed line indicates $T_{c}$.
b) Difference spectrum of the normalised data shown in Fig.~1a.
The solid line represents a Gaussian fit to the data.
}

\label{fig1}
\end{figure}
\begin{figure}
\epsfxsize=3.0in
\epsfbox{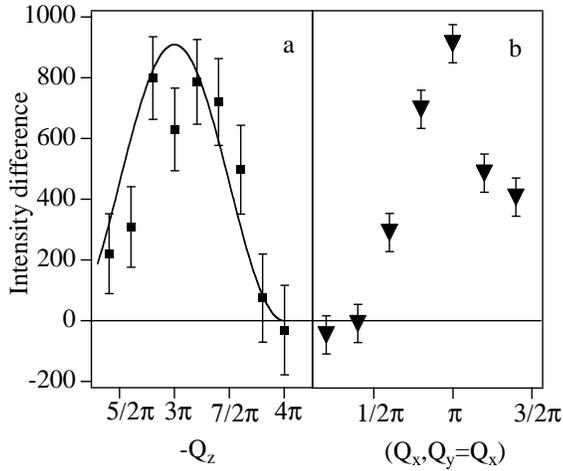}
\vspace{0.5cm}
\caption{a) Difference of neutron intensities
($I(\pi,\pi,Q_{z})-I(0.4\pi,0.4\pi,Q_{z})$)
of Q-Scans perpendicular to the CuO$_{2}$ planes at 34 meV and
$T=10~K$.
The solid line is a sin$^{2}$(Q$_{z}$/2) function describing the odd
character of the resonance along the z-direction.
b) Neutron intensities of Q-scans $I(Q,Q,-3\pi)$ parallel to the
$(\pi,\pi)$ direction at 34 meV, $T=10 K$.
The background measured at $Q_{z}=-2.2\pi$ has been subtracted.}
\end{figure}

\begin{figure}
\epsfxsize=3.0in
\epsfbox{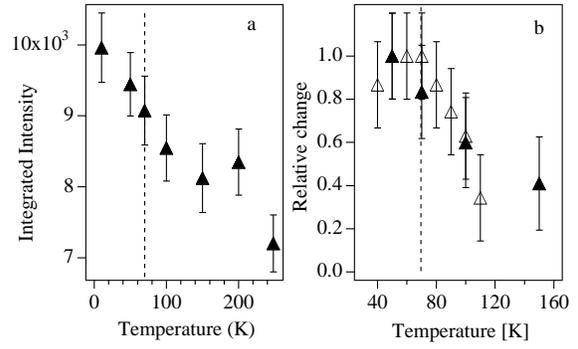}
\vspace{0.5cm}
\caption{ 
a) Temperature dependence of the energy-integrated intensity
of the neutron signal at $(\pi,\pi,-3\pi)$.
b) Temperature dependence of the square of the ARPES gap estimated by the
leading-edge midpoint (open
triangles) and of the neutron intensities (filled triangles). All data
have been normalized to the $T=50~K$ values. For the
neutron data, the high temperature ($T=250~K$) value has been subtracted.
The dashed lines indicate $T_{c}$.
}
\label{fig3}
\end{figure}

\begin{figure}
\epsfxsize=3.0in
\epsfbox{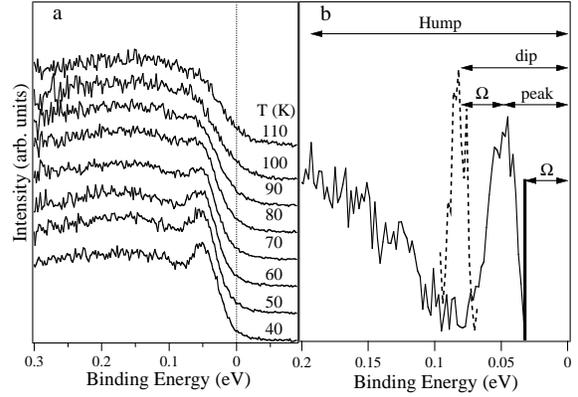}
\vspace{0.5cm}
\caption{
a) Temperature dependence of the ARPES spectra taken at the $(\pi,0)$ point.
b) Expansion of the near E$_{F}$ region at $T=40~K$ together with the
peak, dip, hump and mode energies. Also shown (bold-broken line) is
the magnetic resonance as measured by INS. An offset of 48 meV
(=$\Delta_{0}$),
has been added to the energy axis of the INS data.
}
\label{fig4}
\end{figure}

\begin{figure}
\epsfxsize=3.0in
\epsfbox{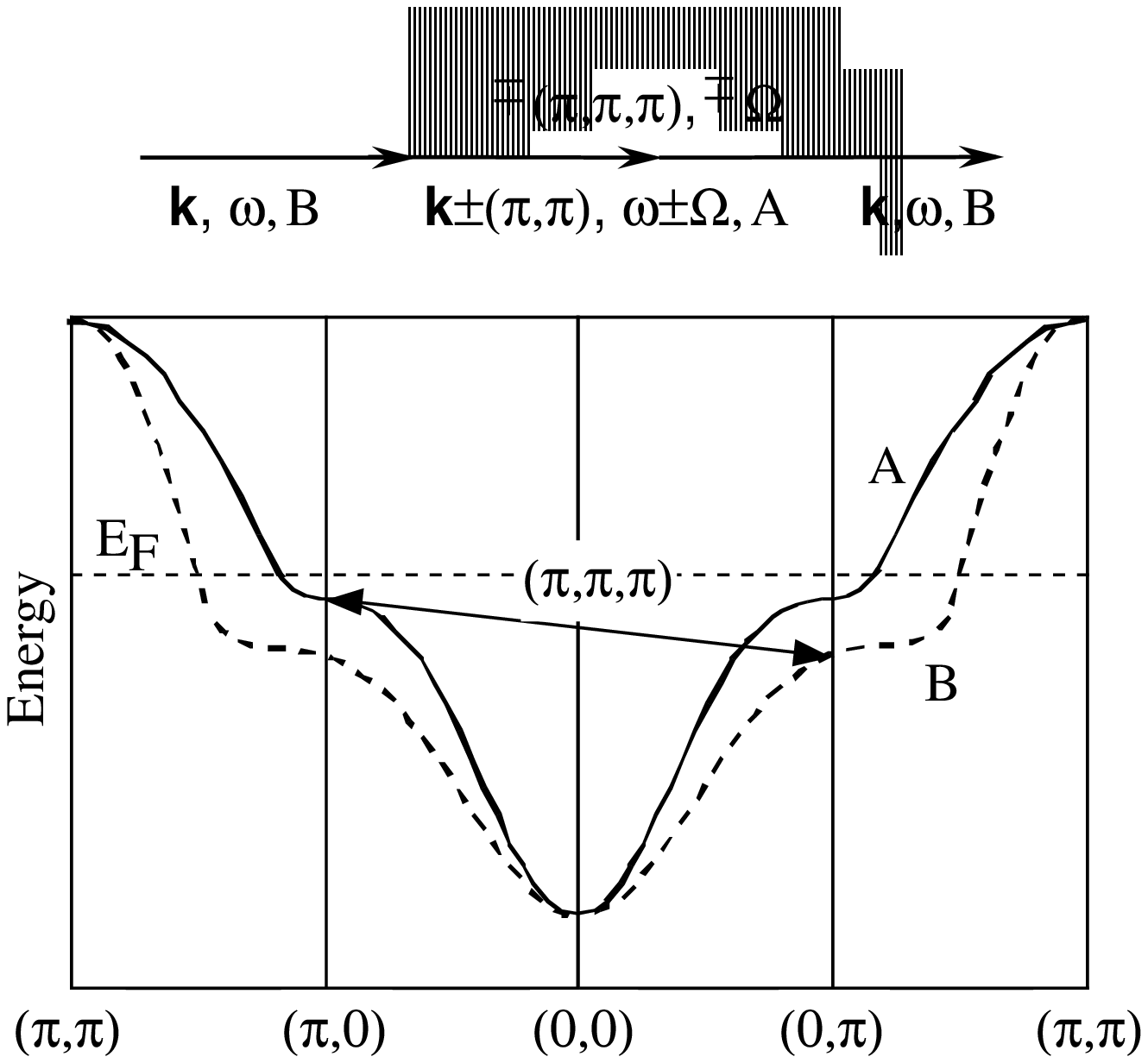}
\vspace{0.5cm}
\caption{
Dispersions of the bonding (solid line B) and
anti-bonding (broken line A) bands assuming a
non-vanishing bilayer splitting.
Also shown are the $(\pi,\pi,\pi)$ wavevector
connecting portions of the Fermi surface around the $(\pi,0)$ and $(0,\pi)$
points in both bands. The interaction diagram of the electron with
the magnetic resonance is shown on the top.
}
\label{fig5}
\end{figure}


\begin{references}
\bibitem{BOSE}
Although the Bose population factor $B(T,E)$ strongly
depends on the energy E, $B(T_{1},E)/B(T_{0},E)$,
is only weakly E-dependent for $20<E<50 meV$ and the temperature range
considered.
\bibitem{NOTE1}
Exploiting the Q-dependence of the Cu-magnetic form factor, we have
obtained similar plots by comparing scans taken at equivalent points
in various Brillouin zones.
\bibitem{FOOTNOTE}
For technical reasons we were
limited to temperatures below 110 K. However, it is known from previous
measurements\protect\cite{JC99} that for underdoped $T_{c}$=70~K,
the pseudogap remains up to a temperature $T^{\star}$ about 300K.
\end{references}
\end{document}